# The refractive index and electronic gap of water and ice increase with increasing pressure

Ding Pan[a,*], Quan Wan[a,b], and Giulia Galli[b]

[a]*Department of Chemistry,*

*University of California,*

*Davis, CA 95616, USA*

[b]*The Institute for Molecular Engineering,*

*University of Chicago,*

*Chicago, IL 60637, USA*

[*]*dpan@ucdavis.edu*

(Dated: April 14, 2014)

## Abstract

Determining the electronic and dielectric properties of water at high pressure and temperature is an essential prerequisite to understand the physical and chemical properties of aqueous environments under supercritical conditions, e.g. in the Earth interior. However optical measurements of compressed ice and water remain challenging and it has been common practice to assume that their band gap is inversely correlated to the measured refractive index, consistent with observations reported for hundreds of materials. Here we report ab initio molecular dynamics and electronic structure calculations showing that both the refractive index and the electronic gap of water and ice increase with pressure, at least up to 30 GPa. Subtle electronic effects, related to the nature of interband transitions and band edge localization under pressure, are responsible for this apparently anomalous behavior.

INTRODUCTION

Water is arguably the most important material in the biosphere, as well as a key constituent of the Earth crust and mantle. Its properties have been extensively studied as a function of temperature (T) and pressure (P) for several decades [1, 2]. However, the electronic structure of water under pressure has remained elusive, due to experimental difficulties in measuring absorption processes in wide band gap insulators in diamond anvil cells. The electronic properties of compressed water are key, for example, to understanding electron charge transfer rates in redox reactions [3, 4] occurring in supercritical water [1], and aqueous solutions in the Earth's interior [2]; such reactions, that depends on both the electronic and static dielectric constant of water, determine the oxidation states of minerals and rocks.

Water was predicted to undergo an insulator-to-metal transition at rather large pressures, estimated between 100 GPa and 5 TPa [6–9], though no experimental confirmation has yet been reported. Experimentally, it is difficult to directly establish the variation of the liquid electronic gap as a function of P, which at ambient conditions is outside the optical window of diamond: the optical gap of diamond is 5.4 eV [10] with the absorption tail as low as $\sim$4 eV, and the quasi particle gap of water at ambient conditions is estimated to be 8.7$\pm$0.5 eV [11, 12] with the optical absorption onset at $\sim$7 eV [13, 14]. Hence the band gap ($E_g$) of high pressure ice VII (a disordered ice phase stable between 3 and $\sim$60 GPa [15, 16]) has been inferred from measurements of its refractive index, $n$, which can be obtained by using low energy photons without probing absorption processes [17]. In particular, it has been common practice to use a single-oscillator model, e.g., the Penn model [18], to relate $n$ and $E_g$, as in the case of other molecular crystals, notably hydrogen[19, 20]. According to the Penn model, the refractive index $n$ and the electronic gap $E_g$ are inversely correlated: $n^2 = 1 + \frac{(\hbar\omega_{fp})^2}{E_g^2}$, where $\omega_{fp}$ is the plasma frequency. This inverse correlation between $n$ and $E_g$ holds for a broad class of materials [21, 22].

Zha et al. measured the refractive index of ice VII as a function of pressure up to 120 GPa and found that it increased with pressure, similar to the case of solid hydrogen. They found weak dispersion of the refractive index, and its pressure dependence, leading them to conclude that the band gap is preserved to very high pressure, ultimately closing only in the terapascal pressures and well beyond the stability field of ice VII. The refractive index of liquid water has been measured only up to 7 GPa, at T = 673 K [23], while no measurements have yet been reported of its electronic gap in such P-T regime.

In order to investigate the electronic and dielectric properties of water under pressure, we carry out ab initio molecular dynamics (MD) simulations and compute, independently, the refractive index $n$ and the electronic gap $E_g$ in the pressure range 0–30 GPa. We obtain values for the refractive index of ice and water in good agreement with experiments [17, 23]; however we find that both $n$ and $E_g$ increase with pressure in the solid and liquid phases. Our results show that a simple single-oscillator model used to rationalize the electronic properties of many semiconductors and insulators, may not be employed to describe water under pressure. We find that due to subtle but important changes in the localization properties of the valence band of water and ice, the strength of inter band transitions is not a constant under pressure. As a consequence $n$ and $E_g$ are not inversely correlated, but exhibit a more complex interdependence, not captured by widely used simple models. Our results are consistent with those of Hermann and Schwerdtfeger who reported an increase of the optical gap of ice under pressure [24], using calculations based on many body perturbation theory with a fitted dielectric constant.

**RESULTS**

**Refractive index of water and ice**

For photon energies ($\hbar\omega$) much smaller than the band gap, absorption processes may be ignored and $n^2 = \epsilon_1$, where $\epsilon_1$ is the real part of the electronic dielectric constant. For simple semiconductors, the Penn model [18] yields a reasonable approximation of $\epsilon_1$. The validity of such model may extend beyond simple semiconductors, as suggested by Wemple and DiDomenico who showed that $[\epsilon_1(0) - 1]$ is approximately proportional to $1/E_g$ for more than 100 solids and liquids[21]. However, as discussed below, the description of the properties of water and ice requires a more sophisticated level of theory.

Fig. 1 reports the calculated refractive index of water and ice, as a function of pressure, and shows that $n$ increases with P along isotherms for both the solid and the liquid. We also show experimental results for ice VII [17] and water up to 7 GPa [23]. To the best of our knowledge our results represent the first data obtained for the refractive index of the supercritical liquid above 7 GPa. Indeed, the highly corrosive character of hot, compressed water makes experiments rather challenging at high P and T.

We found good agreement with experiments, especially above 5 GPa, with the deviations present at low pressure ascribed to errors introduced by the use of the PBE functional [25]. Semi-local functionals have been shown to yield results for water and ice under pressure in better agreement with experiments than at ambient conditions [26], e.g., for the calculation of the equation of state [27] and the dielectric properties [16, 27]. Preliminary calculations of the electronic dielectric constant of water at ambient conditions using the PBE0 [28, 29] functional showed a much improved agreement with experiment, yielding $\epsilon_1 = 1.78$ (to be compared with the experimental value of 1.77). The same improvement was also found for the static dielectric constant of ice Ih[30]. For ice VIII at 30 GPa, we found instead a minor difference of 4% between the electronic dielectric constants obtained by PBE and PBE0.

**Electronic gap of water and ice**

In Fig. 2, we show the electronic gap of water and ice, computed as the difference between the conduction band minimum (CBM) and the valence band maximum (VBM) as a function of pressure. We report values obtained with the PBE functional and, in the inset, with the hybrid functional, PBE0 [28, 29]. In spite of large statistical fluctuations at high T, the trend of an increasing gap with pressure is evident, irrespective of the level of theory used. As well known, the PBE functional underestimates the band gap of water [12, 31], but the gap variation under pressure is qualitatively the same and quantitatively similar within PBE and PBE0. We expect the rate at which the optical gap increases with pressure be larger than for the electronic gap. Indeed, based on our results for $\epsilon_1$, we estimated that the exciton binding energy, $E_b$, of water will be decreased by about 60% in going from ambient pressure to ~30 GPa, and that of ice VII/VIII by 40% (The $E_b$ of water at ambient conditions is about 2.4 eV [31]). Our estimate is based on the relation $E_b = m^*/(m_e \epsilon_1^2 a_0)$, where $m^*$ is the electronic effective mass, $m_e$ the electron mass and $a_0$ the Bohr radius [32], where we assumed a negligible variation of the electronic effective mass under pressure, as indicated by our band structure calculations (not shown). Our results for band gaps are consistent with those reported by Hermann and Schwerdtfeger for ice VIII, using many body perturbation theory [24]. These authors used the $G_0W_0$ approximation and the Bethe-Salpeter equation [24] to compute absorption spectra, with a model dielectric function to approximate the screened Coulomb interaction. In their model, the electronic dielectric constant was an input parameter. Our results are also consistent with those of Boero et al. who reported an increase of the band gap of supercritical water at 653 K, for densities larger than 0.5 g cm$^{-3}$, by using ab initio MD and the BLYP functional [33].

**Relation between refractive index and band gap**

Our calculations showed that both the refractive index and the electronic gap of water and ice increase under pressure, up to at least 30 GPa. They also showed that $n$ of ice is larger than that of the liquid, in spite of the latter having a smaller electronic gap. Thus, the simple inverse correlation between $E_g$ and $n$ used to interpret experiments, and valid for a wide range of substances [21], does not hold for water and ice. We show in the following that two reasons are responsible for this apparent anomalous behavior: (i) an increase of the electronic density with pressure and hence of the plasma frequency $\omega_{fp}^2 = \frac{4\pi \rho_e e^2}{m_e}$, where $\rho_e$ is the density of valence electrons and $e$ the elementary charge; (ii) a change in the localization property of the valence band of the liquid and solid, as P is increased, which in turn is responsible for changes in the strength of interband transitions.

Interestingly, while for ice the increase in $\rho_e$ (and $\omega_{fp}$) counterbalances the increase of $E_g$ and hence $[\epsilon_1(0) - 1]/\rho_e$ decreases under pressure, the corresponding quantity for water shows an increase (see Fig. 3). Ice and water behave as several oxides [34] in exhibiting a positive derivative of the refractive index, with respect to $\rho_e$. However for both water and ice the quantity $[\epsilon_1(0) - 1]E_g^2/\rho_e$ is not a constant, as it would be if the Penn model correctly described the relation between the dielectric constant and electronic gap, but it monotonically increases with pressure, at all T considered here.

We therefore adopted a higher level of theory and derived the real part of the electronic dielectric constant in the random phase approximation (RPA):

$$\epsilon_1(\omega) = 1 + \frac{e^2}{\pi^2 m_e} \sum_{v,c} \int_{BZ} d\vec{k} \frac{f^\mu_{cv}(\vec{k})}{\omega^2_{cv}(\vec{k}) - \omega^2} \quad (1)$$

where $c$ and $v$ are the conduction and valence band indices, respectively, and the integral is over the Brillouin Zone (BZ). In the integrand $f^\mu_{cv}(\vec{k})$ is the oscillator strength in the polarization direction $\vec{e_\mu}$ associated to the transition between the bands $c$ and $v$, and $\hbar\omega_{cv}(\vec{k})$ is the corresponding transition energy. In the case of water, only the $\Gamma$ point was used in our electronic structure calculations; at zero frequency one has :

$$\epsilon_1(0) = 1 + \frac{8\pi e^2}{V m_e} \sum_{v,c} \frac{f^\mu_{cv}}{\omega^2_{cv}} \quad (2)$$

where $V$ is the volume of the simulation cell. Note that in a single-oscillator model with $\hbar\omega_{cv} = E_g$, Eq. 2 reduces to the Penn model, due to the oscillator strength sum rule.

The oscillator strength is given by

$$f^\mu_{cv} = \frac{2 m_e \omega_{cv}}{\hbar} |\langle c | \vec{e_\mu} \cdot \vec{r} | v \rangle|^2 \propto \epsilon_2(\omega_{cv}) V \omega_{cv}, \quad (3)$$

where $\langle c|$ and $|v\rangle$ denote Kohn-Sham orbitals corresponding to valence($v$) and conduction($c$) states, and $\epsilon_2$ is the imaginary part of the dielectric function. For $v$ to $c$ transitions right across the band gap, the oscillator strength is proportional to $[\epsilon_2(\omega_{cv})V\omega_{cv}]$. In Fig. 4 we plotted $\epsilon_2(\omega)V$ of two water configurations at 1000 K, at $\sim$1 GPa and $\sim$10 GPa. These are representative of a set of 25–55 snapshots analyzed for each pressure, extracted from our MD trajectories at equally spaced simulation times; for each of them we found that at $\hbar\omega = E_g$, $\epsilon_2 V$ is much larger at $\sim$10 GPa than at $\sim$1 GPa. The oscillator strength of the transitions just across the band gap, which yields largest contribution to $\epsilon_1(0)$ according to Eq. 2, is thus substantially enhanced when the pressure increases from $\sim$1 GPa to $\sim$10 GPa, and the absorption edge is blue shifted, as expected from our electronic gap calculations.

We conclude that it is because of the enhancement of $f_{cv}^\mu$, which outweighs the increase of $E_g$, that the refractive index of water increases under pressure.

We note that local field effects are not included in Eq. 1 [35]; they are instead taken into account in our density functional perturbation theory (DFPT) calculation of the refractive index [1] [37]. We verified that neglecting such effects amounts to a negligible error in the calculations of $n$. In Fig. 5, we compare $\epsilon_1(0)$ of ice VIII, as well as that of high pressure water, obtained with and without local field corrections. The two sets of values are rigidly shifted by only $0.1 \sim 0.2$ with respect to each other, indicating that none of the trends as a function of pressure reported here is affected by local field effects.

**Localization properties of water and ice valence bands**

In order to understand the changes in the oscillator strength of water under high pressure, we computed the inverse participation ratios (IPRs) of the VBM and CBM orbitals; they are reported in Fig. 6. The IPR is defined as

$$IPR_\alpha = \frac{\frac{1}{N}\sum_{i=1}^{N}|\psi_\alpha(i)|^4}{[\frac{1}{N}\sum_{i=1}^{N}|\psi_\alpha(i)|^2]^2}, \quad (4)$$

where N is the total number of points used to perform integrals over a real space grid, and $\psi_\alpha(i)$ is the wave function associated to the band $\alpha$. The quantity $IPR_\alpha$ yields the statistical variance for the distribution of $|\psi_\alpha|^2$ in the system. The larger the IPR, the more localized the wave function. An $IPR_\alpha$ close to 1 indicates a delocalized electronic orbital. Fig. 6 shows that under pressure, as at ambient conditions, the CBM of water is much more delocalized than its VBM [38]. The interesting finding is that with increasing pressure the localization properties of the CBM are largely unaffected, while those of the VBM change substantially, with a tendency of the VBM to delocalize. The electronic states of ice VIII behave similarly except that the variation of the VBM localization as a function of P is smaller than that of water, as shown in Fig 6(c). As a result, under pressure the spatial overlap between the electron and hole wave functions is enhanced, and the matrix element $\langle c|\vec{e}_\mu \cdot \vec{r}|v\rangle$ increases (see Eq. 3). The fact that the localization of the CBM does not change significantly under pressure, while that of the VBM substantially changes, suggests that the opening of the gap in ice and water under pressure may be mostly due to a change in the position of the VBM.

---

[1] Denote the static dielectric matrix $\boldsymbol{\epsilon}_{GG'}(0)$, where $G$ and $G'$ are reciprocal lattice vectors; $\epsilon_1^{-1}(0) = [\boldsymbol{\epsilon}_{GG'}(0)]^{-1}|_{G=G'=0}$, implying that $\epsilon_1^{-1}(0)$ is the first element of the inverse dielectric matrix [36]. However in Eq. 1, $\epsilon_1(0)$ is simply $\boldsymbol{\epsilon}_{00}(0)$ and so called local field effects are neglected [35].

Experimentally Zha et al. [17] measured $n$ as a function of $\omega$ and obtained two parameters, $E_d$ and $E_0$ from the equation $n^2 - 1 = E_d E_0/(E_0^2 - \hbar^2\omega^2)$. The fit was then used to conclude that the gap of ice decreases under pressure. Note that in the experiment of Ref. [17] when the light wavelength is between 569 to 741 nm, $n$ changes by less than 0.05; as a result, the effective band gap determined from the single-oscillator model changes little with pressure. It is presumably difficult to estimate $E_d$ and $E_0$. The values for $E_d$ and $E_0$ obtained from the fits gave a predicted closing of the band gap only at ultrahigh pressures (∼4.5 TPa), well beyond the pressure range of the measurements. We have shown here that state-of-the-art calculations of the electronic structure are essential to properly interpret complex and sophisticated optical experiments and obtain the correct band gap trend under pressure.

**DISCUSSION**

We reported the first data for the refractive index of supercritical water above 7 GPa and at high temperature. We showed that the refractive index, and the electronic and optical gaps of water and ice increase under pressure, and thus water properties differ substantially from those of apolar molecular fluids and solids, e.g. hydrogen [39], methane [40, 41] and benzene [42, 43], whose electronic gaps decrease with pressure but refractive indexes increase.

We showed that the unusual relationship between $n$ and $E_g$ stems from an enhancement of the strength of interband transitions right across the band gap. This enhancement, in combination with a decrease of the volume, outweighs the increase in the band gap upon compression, and hence the refractive index of water increases with pressure. The behavior of the interband transition strength originates from the delocalization of the valence band edge, whose overlap with the conduction band minimum increases with pressure. Our results showed that the use of simple models, e.g. the Penn model [18], to relate measured refractive indexes to band gaps of molecular fluids and crystals should be revisited and in general complex electronic structure effects should be taken into account, when predicting values of band gaps. We emphasize that the type of electronic structure calculations reported here allows one to infer the band gap of water from high pressure measurements of the refractive index [17]. Measurements of the band gap of water by light absorption or emission are not yet feasible under pressure, in diamond anvil cells; therefore establishing the correlation between the band gap and the refractive index is a crucial step to infer band gaps from measured dielectric constants. In general, ab initio calculations play a key role in interpreting measurements of fluids under pressure, especially in unraveling their electronic structure.

Finally we note that the results reported here for gaps and dielectric constants are expected to have an important impact in the broad field of water-related science. For example, factors determining the rate of electron transfer in aqueous environments (and hence several chemical reactions occurring in these media) include the refractive index and the static dielectric constant of the medium[3]. Hence knowledge of such quantities is key to predict reorganization energies and electron transfer rates in supercritical water and water under pressure, for example in the Earth's interior. In addition, the mobility of charges created in water by, e.g. photo-excitation, depends, among other factors, on the exciton binding energy ($E_b$); our results indicate that $E_b$ decreases under pressure and hence charge mobility is expected to change substantially in supercritical water, with respect to water at ambient conditions, with implications for charge transport under bias, in supercritical conditions.

**METHODS**

**Water**

We carried out ab initio molecular dynamics (MD) simulations of water under pressure as a function of temperature in the Born-Oppenheimer approximation with the Qbox code (version 1.54.2, http://eslab.ucdavis.edu/software/qbox/)[44]. We used 128-molecule unit cells with periodic boundary conditions, plane-wave basis sets and norm conserving pseudopotentials (Gygi, F. Pseudopotential Table, http://fpmd.ucdavis.edu/potentials/index.htm). Atomic trajectories were generated with a plane-wave kinetic energy cutoff of 85 Ry, which was then increased to 220 Ry for the calculation of pressure on selected snapshots [27].

**Ice VIII**

Calculations for ice VIII were conducted by using a 8-molecule unit cell with a plane-wave energy cutoff of 220 Ry and $4 \times 4 \times 3$ k-point mesh, and a 96-molecule supercell with a cutoff of 85 Ry and $\Gamma$ point only.

**Electronic structure calculations**

Trajectories were generated with the Perdew-Burke-Ernzerhof (PBE) exchange correlation functional [25], while electronic gaps and dielectric constants were computed with both the PBE and PBE0 [28, 29] functionals on snapshots extracted from PBE trajectories. At the PBE level, the electronic dielectric constant $\epsilon_1$ was calculated using density functional perturbation theory (DFPT) [37], as recently implemented in the Qbox code for the calculation of Raman spectra [45]. At the PBE0 level of theory, we instead used an electronic enthalpy functional [46]. We calculated the strengths of inter band transitions with the Quantum-expresso package [47]. We used a Gaussian broadening parameter equal to 0.001 eV.

**ACKNOWLEDGMENTS**

We thank F. Gygi, M. Govoni and Y. Ping for many useful discussions and R. Hemley for a critical reading of the manuscript. This work was supported by Department of Energy (DOE), Computer Materials and Chemical Sciences Network, under Grant DE-SC0005180, and by the Sloan Foundation through the Deep Carbon Observatory. Part of this work was carried out using computational resources from the Extreme Science and Engineering Discovery Environment (XSEDE), provided by the National Institute for Computational Sciences (NICS) under Grant TG-MCA06N063, which is funded by the National Science Foundation; and resources at Lawrence Berkeley National Laboratory, NERSC, supported by DOE under Contract No. DE-FG02-06ER46262.

**CONTRIBUTIONS**

D. P. and G. G. designed the research. Most of the calculations were performed by D. P., with contributions from all authors. Q. W. implemented Density Functional Perturbation Theory (used for the calculation of the dielectric constants) into the Qbox code. All authors contributed to the analysis and discussion of the data and the writing of the manuscript.

**COMPETING FINANCIAL INTERESTS**

The authors declare no competing financial interests.

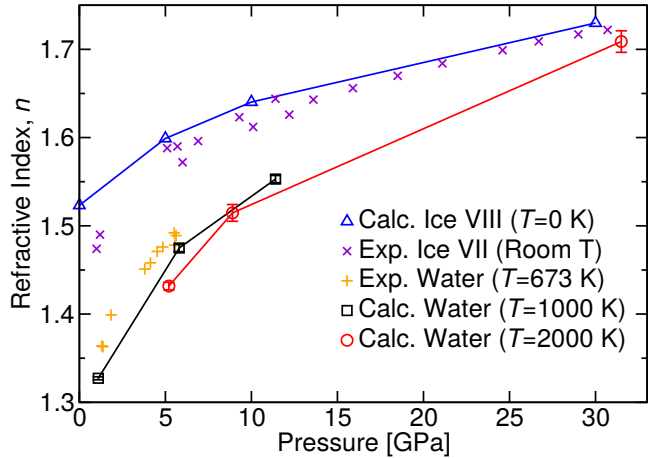

FIG. 1. **Refractive indexes.** Pressure dependence of the refractive index, $n$, of water and ice. We computed $n = \sqrt{\epsilon_1}$, where $\epsilon_1$ is the real part of the electronic dielectric constant, obtained from ab initio simulations (see text). The experimental data for ice VII, obtained with light wavelength of 630 nm, are from Ref. [17]. The experimental values for water at 673 K are from Ref. [23]. We note that the statistical fluctuations in the value of $n$ are rather small, indicating that $\epsilon_1$ is largely insensitive to reorientations of the molecules in the fluid and to hydrogen bonding rearrangements, unlike the ionic dielectric constant [27]. When increasing the temperature from 1000 to 2000 K, $n$ showed again moderate variations. The error bars show the standard derivations of the refractive indexes of water. Lines are drawn to guide the eyes only. Calculations were conducted for ice VIII instead of VII for computational simplicity, as ice VII is proton disordered.

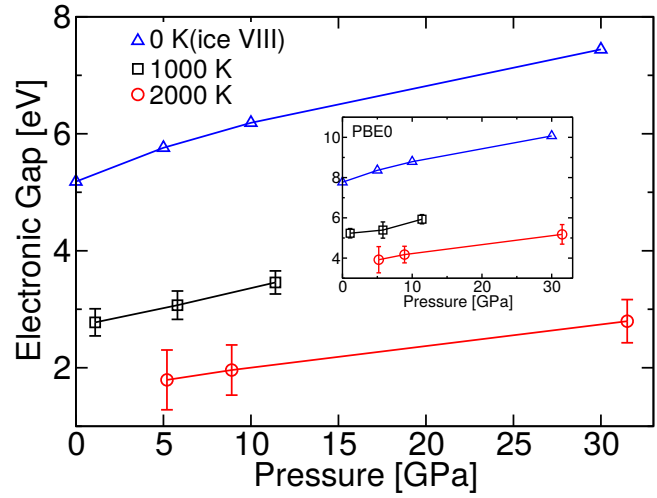

FIG. 2. **Electronic gaps.** Computed electronic gap of water and ice VIII as a function of pressure at 0, 1000 and 2000 K. The calculations were performed using the semilocal density functional, PBE, and in the inset, the hybrid functional PBE0. The error bars show the standard derivations of the band gaps of water. Lines are drawn to guide the eyes only.

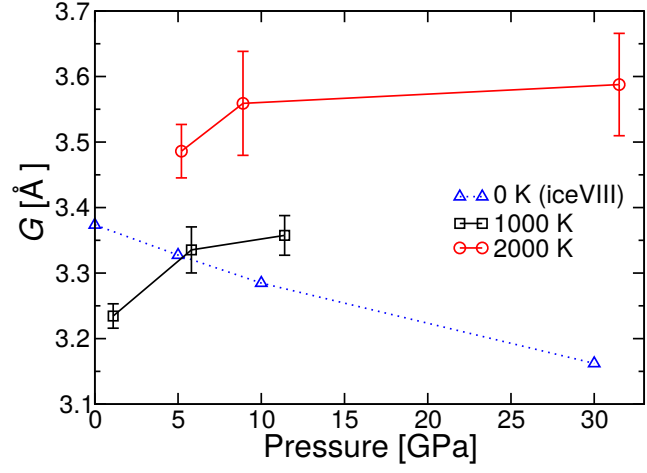

FIG. 3. **Volume effect on refractive indexes.** This figure shows $G = (\epsilon_1(0) - 1)/\rho_e$ as a function of pressure at 0, 1000 and 2000 K, where $\epsilon_1(0)$ is the real part of the electronic dielectric constant at zero frequency and $\rho_e$ the valence electron density. The error bars show the standard derivations of the data for water. Lines are drawn to guide the eyes only.

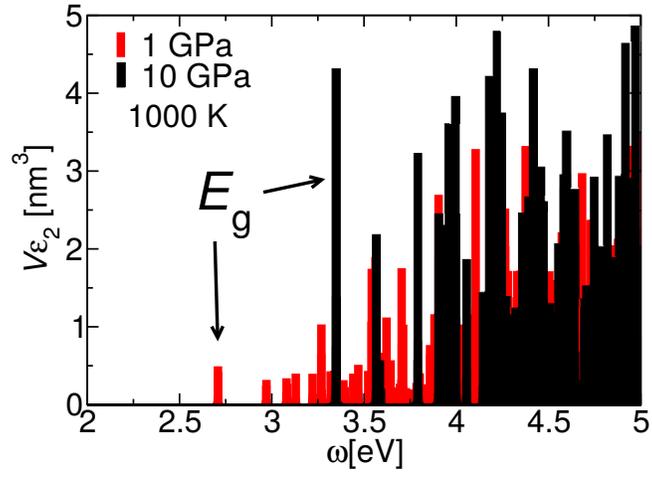

FIG. 4. **Oscillator strength of interband transitions.** The imaginary part of the electronic dielectric constant of water, $\epsilon_2$ multiplied by the cell volume $V$ as a function of frequency $\omega$ at ∼1 and ∼10 GPa and 1000 K. $E_g$ labels the frequencies corresponding to the band gaps of water at ∼1 and ∼10 GPa, respectively.

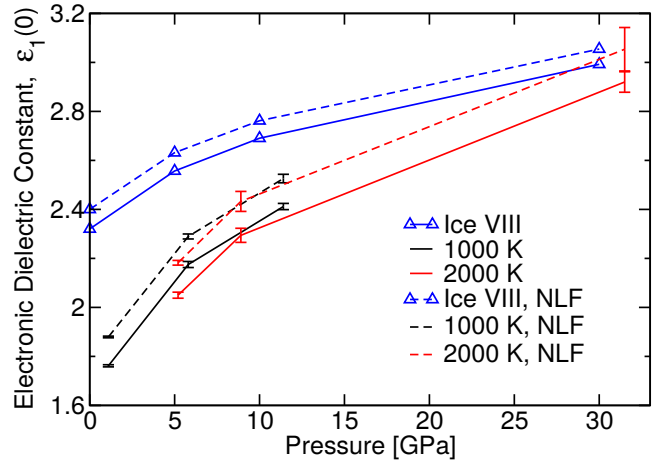

FIG. 5. **Local field effects on dielectric constants.** The dashed lines show the real part of the electronic dielectric constants of water and ice VIII (0 K) at zero frequency, $\epsilon_1(0)$, obtained from density functional perturbation theory (DFPT) without local field effects (NLF). As a comparison, the results obtained with DFPT and including local field effects (as done in Fig. 1), are shown by the solid lines. The error bars show the standard derivations of the electronic dielectric constants of water

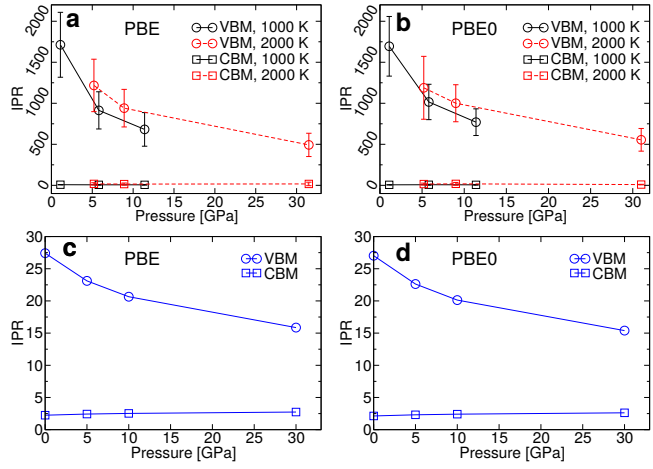

FIG. 6. **Localization properties of water conduction and valence bands.** The inverse participation ratios (IPR) of the valence band maximum (VBM, circles) and conduction band minimum (CBM, squares) of (a, b) water and (c, d) ice VIII as a function of pressure. Results of panels b and d were obtained with the hybrid functional PBE0, while those of panels a and c with the semi-local functional PBE. The error bars show the standard derivations of the IPRs of water. Lines are drawn to guide the eyes only.